# Deep learning for the detection of machining vibration chatter

[1,2] **Cheick Abdoul Kadir A. KOUNTA,** [1]**Lionel ARNAUD,** [1]**Bernard KAMSU FOGUEM,** [2]**Fana TANGARA**

[1]Laboratoire Génie de Production,

École Nationale d'Ingénieurs de Tarbes

47 Avenue Azereix, BP 1629, F-65016 Tarbes Cedex, France

[2]Faculté des Sciences et Techniques

Université des Sciences, des Techniques et des Technologies de Bamako (USTTB)

B.P. E : 423, Bamako, Mali

Cheick Abdoul Kadir A. KOUNTA (ckounta@enit.fr), Lionel ARNAUD (lionel.arnaud@enit.fr ), Bernard KAMSU-FOGUEM (bernard.kamsu-foguem@enit.fr), Fana TANGARA (fbtangara@yahoo.fr)

**Highlights**

1. Detection of chatter in industrial machining using Fast Fourier Transform and deep learning.
2. The dataset is obtained via transformations on the vibratory signal of the machining process with the help of a professional expert.
3. The proposed approach uses pre-trained models for feature extraction and classification.
4. The model trained on the unambiguous signal can be generalized by classifying ambiguous and unambiguous signals.
5. The model has been trained on renormalized signals, showing its ability to detect chatter without amplitude information.

**Abstract**

Most of the work on chatter detection is based on laboratory machining tests, thus without the constraints of noise, the variety of situations to be managed in the industry, and the uncertainties on the parameters (sensor position, tool engagement, and sometimes even spindle rotation frequency). This work presents an approach first based on mechanical skills first to identify the optimal signal processing, then based on deep learning to automatically detect the phenomenon of chatter in machining. To do so, we use input images from the Fast Fourier Transform analysis of a vibration signal from the machining of an Alstom industrial railway system. The FFT images are first cropped in height to perform a renormalization of the signal scale. This renormalization step eliminates a bias regularly observed in the literature, which did not eliminate a simple correlation with signal amplitude because chatter is often, but not always, associated with stronger vibrations. Deep learning extracted features from the image data by exploiting pre-trained deep neural networks such as VGG16 and ResNet50. The data is derived from an industrial vibration signal (with noise, a not ideally placed sensors, and uncertainties on process parameters), which was very poorly represented in the literature. It is shown that by depleting the signal (renormalized amplitude and absence of information on the rotation speed), the model obtains a good result (an accuracy of 99% in the training and validation phases), which had never been done in the literature. In addition, since a part of the test dataset is similar to the training dataset, the proposed model obtains an accuracy of 98.82%. The robustness is proved by adding another test dataset containing noisy and more or less ambiguous images to be classified. The model still manages to detect these cases from its training of unambiguous cases, showing that the obtained generalization is efficient with an accuracy of 73.71%.

**Keywords**: deep learning; vibration signal analysis; chatter; fault detection; Fourier transform; machining.

# 1. Introduction

Chatter, described by several historical authors (Taylor, 1907; Tobias and Fishwick, 1958), corresponds to the interaction between a part and a cutting tool in a machining process. Chatter results in irregular marks (defects) on a machined surface. It represents one of the most complex difficulties in typical machining processes, such as turning, milling, drilling, and grinding.

Vibration measurements are widely applied in machine monitoring, allowing quick and accurate detection of rotating machine malfunctions (Manhertz and Bereczky, 2021). Vibration analysis of signals can provide status indicators of existing vibration phenomena, help ensure the quality of machined parts, and facilitate decision-making to remedy problems. Analysis techniques can detect weak signals as early as possible by establishing real trend curves (Vishwakarma et al., 2017). A vibration spectrum is data in which it is possible to read extremely tenuous defects, such as the chipping of a ball in a bearing. Vibration analysis of signals is one of the most popular ways to monitor the condition of rotating machines in operation (Nirwan and Ramani, 2021).

The identification, stability, and detection of chatter have been growing research areas for decades. Recent research on machining chatter detection focuses on advanced applications of machine learning or deep learning due to their ability to solve highly complex and non-linear problems related to high-dimensional datasets (Goodfellow et al., 2017; LeCun et al., 2015). Artificial Intelligence (AI) applied to vibration analysis would be helpful for both experts and non-specialists. For the former, it would avoid wasting time analyzing signals from machines in good condition or with apparent defects. Experts could monitor a more extensive fleet and spend more time on complex cases to reduce the risk of error in the analysis. For non-experts, artificial intelligence would allow them to propose a relatively reliable first diagnosis, giving them new autonomy in their decision-making and optimizing the call to experts.

Over time, research has focused on AI applied to the industrial domain, and many studies show (Angelopoulos et al., 2019; Lee and Lim, 2021; Usuga Cadavid et al., 2020) that these techniques are among the main catalysts for moving a traditional manufacturing system to Industry 4.0. With the digital development of the manufacturing industry, manufacturing processes have become more diverse, automated, and customized. The multiple requirements for manufacturing parts have increased significantly. Anomalies in the machining process must be systematically detected, whether they are easily observable, such as in finishing operations, or not, such as in roughing operations, which are usually not subject to systematic observation. The productivity of any milling operation relies on the ability to remove the material as quickly as possible (Arriaza et al., 2018). However, the material removal rate is reduced due to chatter and all the accompanying side effects, such as poor surface finish, tool damage, spindle damage, and noise, for example. Researchers have developed several methodologies to control chatter (Yan and Sun, 2021), but this phenomenon still limits the productivity of most machining operations.

Most existing chatter detection methods are based on chatter indicators that extract features from the signal, usually a simple scalar number. Still, the thresholds of the chatter indicators designed in this way must usually be identified manually for each cutting condition under consideration. The data-based automatic chatter detection problem can be considered a classification case in machine learning (El-Sappagh et al., 2020).

In this paper, we seek to demonstrate that deep learning is capable of detecting chatter on an industrial signal, considering that the amplitude of the signal and the rotation frequency are in fact, not reliable information. The method is composed of two types of transformation: a transformation on the physics (from the time domain to the frequency domain) and a transformation on the data by exploiting several pre-trained networks, VGG16 and ResNet50. This distinguishes the proposed model from classical monitoring methods that are generally based on signal amplitude and frequency peak analysis while knowing the rotation speeds very precisely.

Most of the methods, probably all of them, explicitly or not, are based on analyzing the amplitude of the vibration signal and/or comparing the measured frequencies with the spindle rotation frequency. However, in laboratory tests, the correlation between the chatter phenomenon, the vibration amplitude, and the frequencies that appear is most of the time trivial, raising questions about the interest in the sophisticated detection methods proposed.

In an industrial situation, this amplitude and frequency information is only sometimes very reliable due to the highly variable position of the sensors and the fact that secondary maintenance operations often create a discrepancy between the real and assumed rotation speed. Thus, to determine if our approach would be of real interest in an industrial situation, we chose to erase the information on the spindle speed and the average amplitude of the signal. Then, we use transfer learning methods to extract the features. These methods extract the features from the raw signal image data to detect machining phases related to vibration phenomena during the machining process: chatter, machining without chatter, and rotation without machining.

The rest of this paper is organized as follows. Section 2 provides the state of the art of deep learning applications for monitoring and detecting faults in machining processes. Section 3 discusses the proposed approach for the detection of chatter in machining. Section 4 presents the chatter detection results of the proposed method. The discussion and the conclusion are presented in Section 5.

## 2. State of the Art

In the literature, two groups of authors have been identified in the study of the chatter phenomenon (Kounta et al., 2022). The first group of authors publishing historically on machining and, in particular, machining chatter (J. Tlusty, Y. Altintas, J. Munoa, etc.), and authors publishing on sensors (H.O. Unver, B. Sener, etc.) who have developed others chatter detection techniques.

Several works in the literature seek to detect chatter with signal processing techniques (FFT, STFT, WT, EMD, HHT, etc.) and have shown some effectiveness for the task (Xu et al., 2006). The authors (Huda et al., 2020) experimented, for example, with a turning process, simply correlating the chatter to the amplitude signal and specific peaks in FFT. They collect two types of signals (normal and chatter) with a microphone very near the machining zone. (Ding et al., 2010) measured the cutting force using a dynamometer table in the machine and applied an FFT to identify the chatter by comparing the power spectrum with a predefined peak threshold. (Chen et al., 2019) propose an approach to extracting the chattering characteristics from an accelerometer sensor signal (put on a dynamometer table), by analyzing the images of the dominant frequency bands through the spectrograms from the STFT. (Afazov and Scrimieri, 2020), as many other studies, detect chatter with the FFT on the measured cutting forces. The wavelet transform (WT) has also been applied to the cutting force measured by a dynamometer table, to detect chatter (Yoon and Chin, 2005). Some authors also use analytical and numerical methods, such as the stability lobe theory, to better identify chatter frequencies (Altintaş and Budak, 1995; Smith and Tlusty, 1993).

These analysis techniques have always had difficulty detecting chatter at an early stage, i.e., before the amplitude of the vibrations is already significant, and visible on the amplitude of the signal, and before it is too late to preserve the quality of the part.

The chatter is mathematically described by strongly non-linear equations, mainly because of a delay term in differential equations of the mass-spring type, which are not easy to master from a mathematical or numerical point of view. Signal processing tools or analytical methods usually relies on identifying a threshold that characterizes the occurrence of chatter. This task, traditionally performed by an expert, can be facilitated by machine learning methods. Machine learning methods SVM and ANN (HongQi et al., 2011; Yao et al., 2010) are commonly applied in vibration signal analysis. An intelligent multi-sensor monitoring system with three units for turning processes is proposed by (Zhang and Shin, 2018). The three units are implemented by integrating two low-cost sensors, an energy meter, and an accelerometer. This integrated monitoring device provides high prediction accuracy for all three process conditions, generalized applicability over various operating conditions, and satisfactory robustness to process and measurement uncertainties. For chatter detection, they achieve 96.88% accuracy even when changing the configuration. (Chen and Zheng, 2018) and (Zheng et al., 2022) propose a chatter detection method in end milling based on wavelet packet transform (WPT) and SVM as feature extractors. (L. Zhu et al., 2020) decompose the raw cutting force signal by the EEMD function to obtain a set of Intrinsic Mode Functions (IFMs). The principal component analysis method retained the IFMs containing chatter situations and underwent a dimension reduction process. A non-linear SVM then performs the detection process. (Shrivastava and Singh, 2019) propose another machine learning method, by the artificial neural network, to predict tool chatter in a turning process. (Khasawneh et al., 2018) combine supervised machine learning with topological data analysis (TDA) to produce a process descriptor capable of detecting chatter.

These methods are developed in three steps: data collection, feature extraction, and model training. The decision phase then requires no human intervention, but there are some limitations, such as sensitivity to measurement errors and expert labeling of training data. These limitations can be addressed by unsupervised learning techniques such as the K-means method (Dun et al., 2021) which does not require labeling, or by deep learning methods.

For the design of methods capable of detecting abnormal behavior on machines, deep learning allows the development of new diagnostic techniques for monitoring the vibration signal. Deep learning aims to develop algorithms allowing modeling with a high level of data abstraction via architectures with cascades of linear and non-linear operations. This learning is notably related to classification (discrete outputs) or regression (continuous outputs) to predict the answers to questions asked on extensive datasets (LeCun et al., 2015). Deep learning has become a popular technique for extracting information (features) from data due to its ability to process raw data and automatically recognize features across multiple levels of abstraction (Glaeser et al., 2021).

Integrating deep learning into intelligent fault diagnosis of rotating machines has achieved great success and benefited industrial applications (Zhang et al., 2021). Compared to conventional machine learning, deep learning automatically extracts features at a higher level and merges feature extraction and classification into a single structure so that a large amount of trial and error is not required, resulting in reliable performance (Janiesch et al., 2021; Wang et al., 2021). Problem anticipation in the industrial context is paramount for identifying weak signals and reducing response times. The objectives are multiple: (i) the reduction of the number of breakdowns, (ii) the reliability of the production tool, (iii) the increase of the availability rate, and (iv) the improvement of the spare parts stock management. Indeed, from the vibrations regularly collected on a rotating machine, the vibration analysis consists of detecting or anticipating possible malfunctions (Glowacz et al., 2019).

Detecting these malfunctions will often be done via image processing which benefits from a set of detection and analysis methods to allow the automation of a specific task. This technology can classify objects or detect an anomaly or a manufacturing defect in an image. Segmentation also represents a detection task to locate an element on an image to the nearest pixel. Image analysis thus offers powerful means to visualize and recognize the defects identified on the systems considered.

According to (Hoang and Kang, 2019), intelligent defect detection comprises three steps: data collection, feature extraction, and defect classification. Vibration sensors are widely used for data collection due to their availability and high sensitivities. This process depends on the data that reflects the problem at hand. There are several tools to transform the signals obtained from these sensors, such as Fourier transform (Bengherbia et al., 2020; Lin et al., 2016), Mortelet wavelets (Behera et al., 2021), and Daubechies, Mallat, and Meyer wavelet packet transform (Xiong et al., 2020). The feature extraction process is essential for obtaining good accuracy in intelligent sensing. This process is divided into two types, physics-based and data-based methods. Physics-based methods usually extract features from the time, frequency, and time-frequency domains. Data-driven strategies use machine or deep learning techniques to extract the most informative features from the data available (any sensors, image, or text information).

(Liao et al., 2021) propose a method for monitoring the manufacturing process through the fusion of time-frequency analysis and deep neural networks. Acoustic emission signals were acquired during turning operations with different spindle speeds, feed rates, and depths of cut. It can be seen in the literature review that the authors address problems such as process and machine condition monitoring in real time, detection and diagnosis of bearing faults in rotating machines, and chatter detection by taking into account aspects such as the change in rotational speed, feed rate, and very generally the average band frequency amplitude.

A hybrid model that combines a pre-trained deep neural network Alexnet (Krizhevsky et al., 2012) and a model based on self-excited vibration theory, based on short-term Fourier transform STFT has been used by (Unver and Sener, 2021) to detect chatter. (Sener et al., 2021) apply the continuous wavelet transform for signal preprocessing by obtaining images rich in chatter information. These images are then exploited as training data for a deep neural network. (Tran et al., 2020) present an approach for real-time chatter detection. This approach is based on the continuous wavelet transform scalogram and convolutional neural networks to predict states (stable, transient, and unstable). A transfer learning model composed of analytical solutions and a convolutional neural network (CNN) is proposed by (Unver and Sener, 2021) to detect the chatter without the need for real data for the training phase, but trivially based on an RMS threshold. (Shi et al., 2020) use a variant of LSTM to detect chatter in high-speed milling processes. On-line chatter detection based on the current signal from a CNC machine is presented by (Vashisht and Peng, 2020). An LSTM is then trained to detect chatter found on the simulated sequence of control currents.

In the literature, many deep learning techniques have obtained satisfactory results. These studies generally use data from machining conditions far from industrial applications, but some papers use realistic cutting conditions.

In this paper, we develop a machine learning method that uses vibration analysis, with specific processing of FFT images, to eliminate possible biases on the known amplitude and frequencies. Then deep learning is used for machining chatter detection, with an additional verification phase on a very different dataset, so-called ambiguous signals, not used during the training.

## 3. The proposed Approach

The proposed process in this work contains multiple steps Figure 1, which are data acquisition, foundation building, deep learning training, validation, and testing of the proposed deep learning model, as well as the extraction of features and construction of the chatter detection model in machining. All these steps will be detailed in the following subsections.

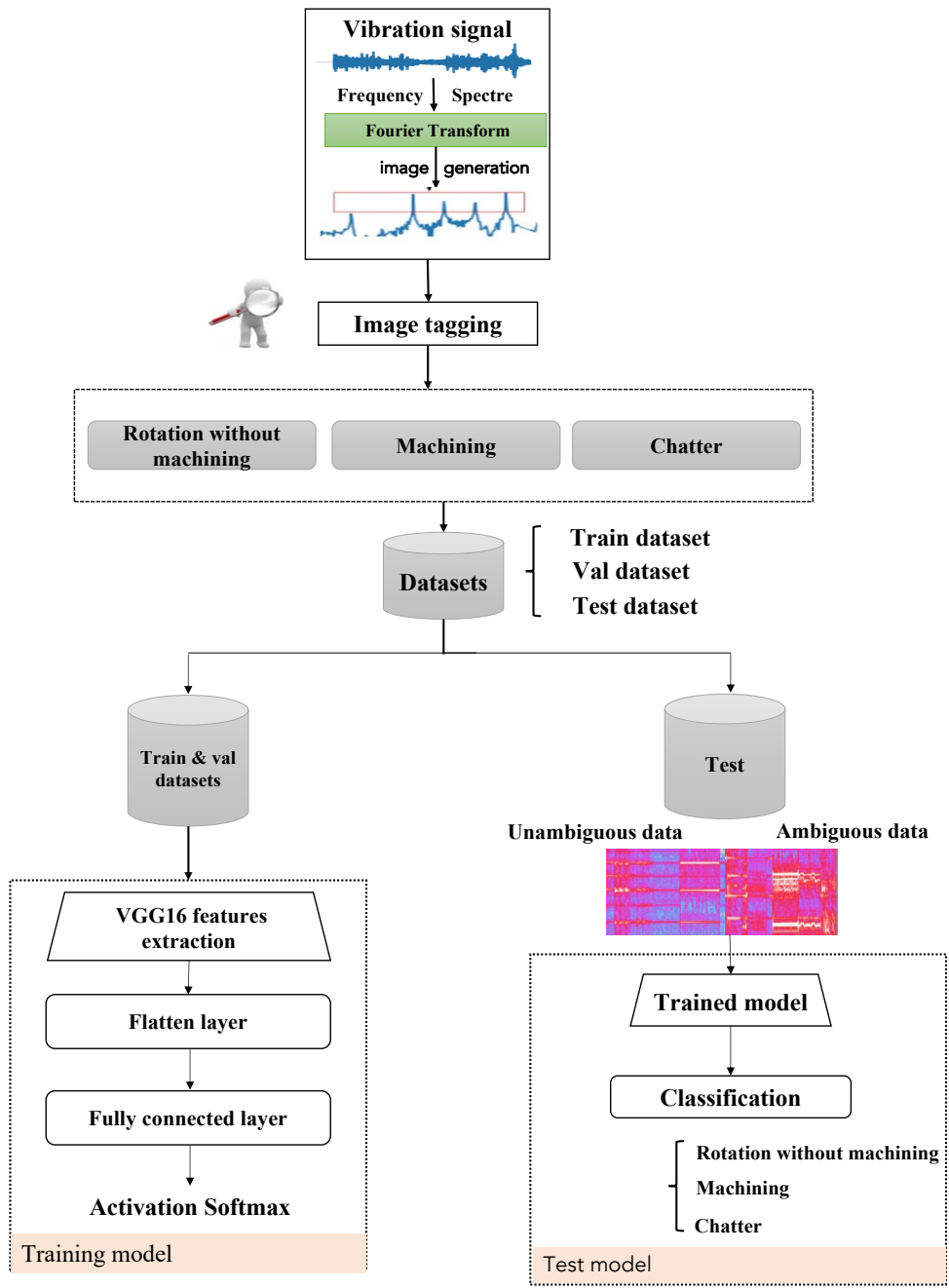

*Figure 1. Proposed process for chatter detection in machining.*

The data acquisition step allows for obtaining a temporal signal from the machining process of train car walls of the High-Speed Train (TGV) company Alstom Transport. This step is followed by transforming from the time domain to the frequency domain to generate FFTs images. Then, these generated images are labeled into three categories to build the training dataset, a validation dataset, and a test base that includes similar data but also noisy and ambiguous signals not used before. Image features are extracted from the dataset by a pre-trained model to train the classification or chatter detection model. The last step will evaluate and verify the model's generalization capability on the two test dataset.

### 3.1. Data acquisition

In machining vibration analysis, the time signal is a crucial element. It represents the information transmitted by the machine tool's sensors from the stresses generated by the mechanical manufacturing process. As the first step in any measurement, vibration analyzers record this time signal and then process it to extract various characteristics. Generally, the size of the time signal for analysis depends fundamentally on the sampling frequency and the period considered. A wrong setting of these parameters can easily lead to a signal that does not reveal any anomaly information, despite the existence of defects. In this case study, the time signal was recorded using a Kistler accelerometer type 8776A50M6, magnetized to the workpiece holder, connected to a Roga Plug-n-Daq signal conditioner, and recorded at 22 kHz using Audacity software. The measurements were made during a machining process of Alstom Transport for the milling of TGV train car walls, illustrated in Figure 2.

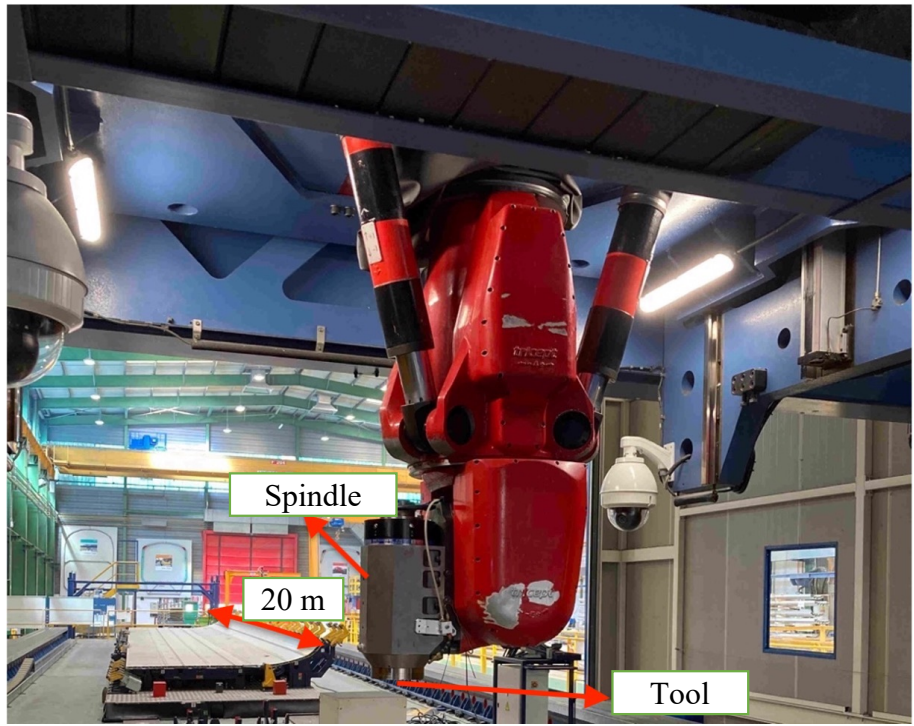

*Figure 2. Milling process of a TGV train car wall (length ≈20 meters). (Image from the machining center of Alstom industry)*

After analysis of the signal with the help of an expert in machining vibration analysis, several machining phases were identified: rotation without machining, machining without chatter, and machining with chatter. The expert made precise identification and labeled these distinct phases of the signal in a spreadsheet file containing the period and label of each phase. We then exported the signal in .wav format and renormalized the signal. An example of the temporal signal used for data collection is visualized in Figure 3, which shows a substantial variation of amplitudes that can be due to the chatter's appearance or to the sensor proximity of the machining zone throughout the many hours of machining.

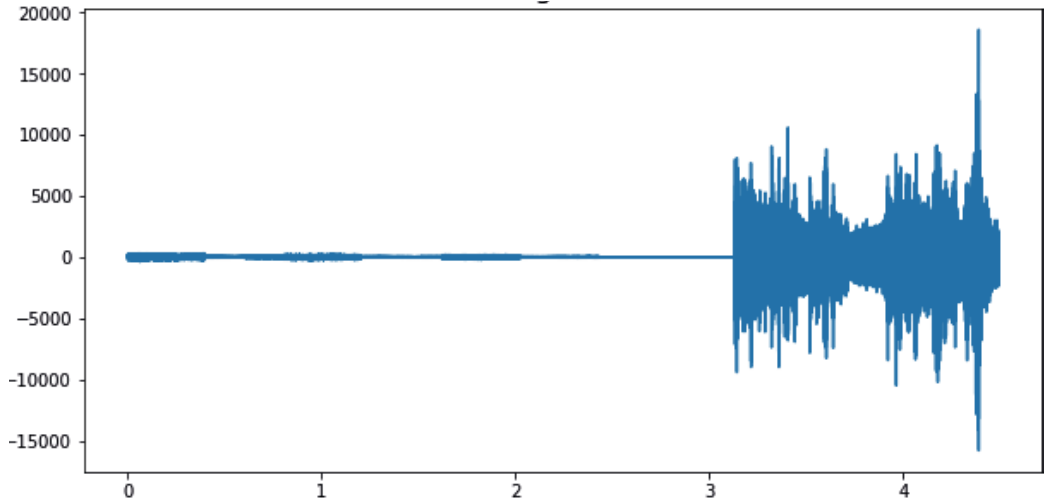

*Figure 3. Temporal signal (time in seconds).*

### 3.2. Preparation of the data

The temporal signal in the .wav file is processed with the Python language using the Librosa library to process the signal. The most common transformation of frequency domain analysis used by machining vibration experts is the Fast Fourier transform (FFT), which can be used to obtain narrowband spectrum efficiently. In agreement with the expert, several FFT was performed in the labeled intervals every 0.1 seconds to optimize the richness and the number of labeled data. Then, images of the FFTs were generated. To reduce the size of the image and focus on the useful part of the signal, the frequency is sampled on 1024 lines in values between 0-2500Hz. The decibel range is set between the maximum and a reduction of 20 dB. This reduction allows us to keep only the upper part of the FFT image, i.e., the dominant peaks, which is the most informative (see the red area in Figure 4). An overview of this transformation is presented in Figure 4.

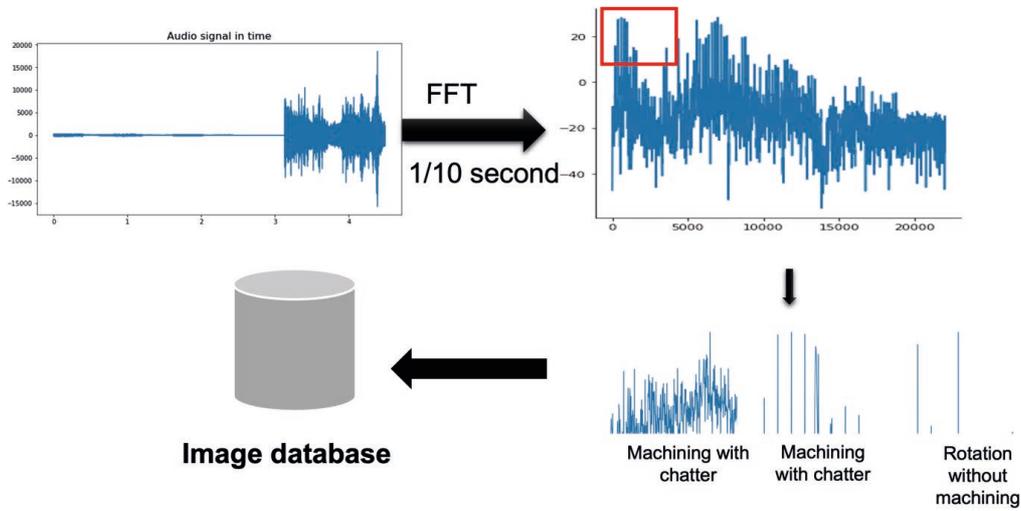

*Figure 4. The transformation from the time domain to the frequency domain.*

The red area in Figure 4 can be seen in Figure 5. These images are then sorted according to their respective labels (rotation without machining: 1582 images, machining without chattering: 5514 images, and chattering: 3089 images) to build the training (representing 70% of the images, i.e., 7126 images) and validation (representing 30% of the images, i.e., 3054 images) dataset.

*Table 1. Distribution of the dataset.*

| Samples | Training (non-ambiguous signals) | Validation (non-ambiguous signals) | Test (non-ambiguous signals) | Test 2 (ambiguous signals) |
|---|---|---|---|---|
| chatter | 2161 | 926 | 1544 | 104 |
| Machining without chatter | 3859 | 1654 | 2814 | 216 |
| Rotation without machining | 1106 | 474 | 733 | 9 |
| **Total** | **7126** | **3054** | **5091** | **329** |

The test is performed on two datasets, a test dataset (5091 samples) containing signals similar to those used in the training and validation phases. Another type of signal was used, which is much more difficult for a human expert to analyze, typically 2x, 3x, or more time to analyze, and are so-called ambiguous signals (329 samples). The difference between the samples in these two datasets can be visualized in Figure 5.

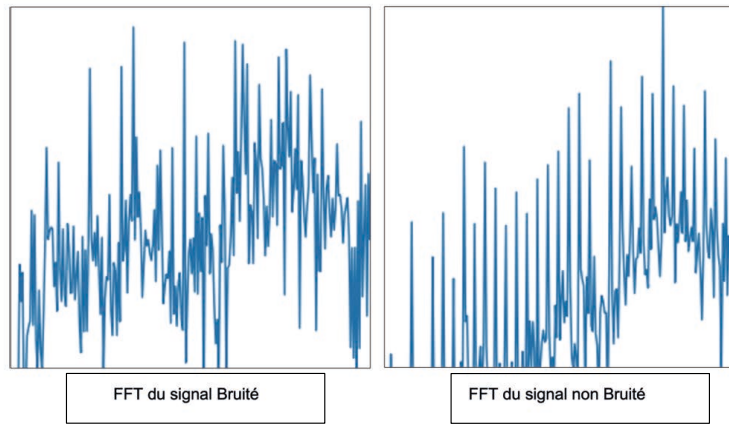
*Figure 5. Images of the ambiguous and unambiguous signal.*

These signals are not noisy in the sense of random noise. Still, they are signals with intermediate characteristics, which require the expert to perform much more in-depth analyses to identify the dominant phenomenon. Analyzing such signals could save the expert much time for signal analysis and the labeling phase. These data will be used to test the ability of the proposed model to classify data that it has never seen and would have been difficult to label.

### 3.3. Transfer learning

The transfer learning technique is widely used in deep learning. It applies knowledge already learned from other models to solve similar problems. Several model architectures can be used in a transfer learning approach. Transfer learning uses existing models that have been carefully selected for their effectiveness and designed by experts to achieve high performance in the target task under consideration, like image classification. Since deep neural networks require a large amount of data, transfer learning is used to learn how to map from an essential characteristic input valid for problem-solving to the desired output. Transfer learning is a fast and efficient way to deal with qualitative or quantitative data shortages.

Implementing the deep neural network model LeNet by (LeCun et al., 1989) sparked the possibility of using convolutional neural networks in practice. AlexNet, invented by (Krizhevsky et al., 2017), is one of the models that proved that CNN could work well on this object recognition dataset (ImageNet). The AlexNet has achieved this complex identification task with good performance, which led the company into a CNN development competition, the ImageNet large-scale visual recognition challenge (ILSVRC). There are models like VGG (Simonyan and Zisserman, 2015), Inception (Szegedy et al., 2016), ResNet (He et al., 2016), to EfficientNet (Tan and Le, 2019). These architectures have almost the same composition: combinations of convolution layers and clustering layers followed by fully connected layers, often with layers having an activation function and/or a normalization phase. In this paper, two pre-trained neural network models are chosen (VGG16 and ResNet50) to extract features from the data. These models have won the Imagenet Large-Scale Visual Recognition Challenge (ILSVRC) at least once.

VGG16 is a convolutional neural network architecture with thirteen convolution layers, five Max Pooling layers, and three Dense layers (fully connected layers).

VGG16 (Figure 6) is a large convolutional neural network model with 138 million parameters widely used on ImangeNet.

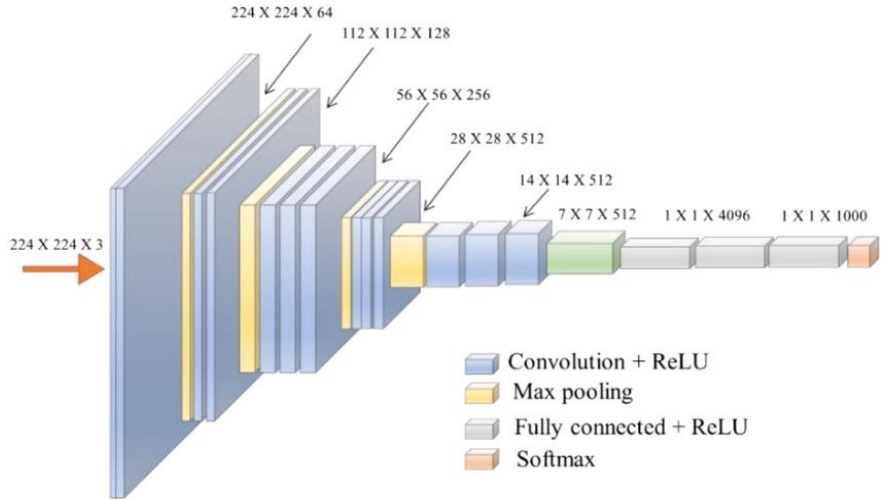

*Figure 6. The architecture of the VGG16 model was proposed by (Pandiyan et al., 2019).*

ResNet50 is one of the most popular convolutional neural network architectures. It is a 50-layer deep neural network model with 48 convolution layers, a MaxPooling layer, and an Average Pooling layer Figure 7. The ResNet50 solved the problem of overlearning and accuracy saturation due to increasing network depth. This model uses two kinds of residual bottleneck blocks (Identity Block and Convolution Block) to reduce the number of parameters and matrix multiplications.

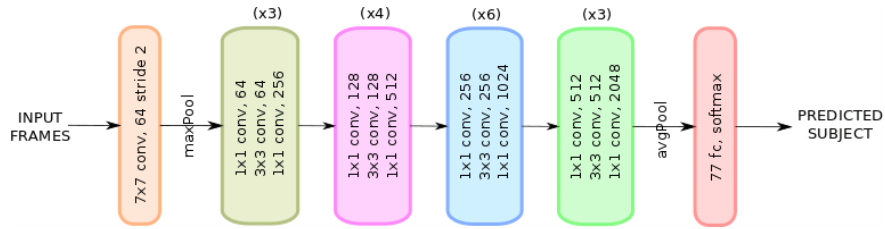

*Figure 7. Architecture of ResNet50 model proposed by (S. Jahromi et al., 2019).*

These pre-trained models are fitted to the dataset for vibration chatter detection in machining. In deep learning, a convolutional neural network comprises two major parts, a lower part to extract features useful for image recognition and an upper part for the classification task. Only the bottom portion for feature extraction is considered, and the classification part is adapted to the dataset of this case study since the VGG16 model, for example, has 1000 classes.

## 3.4. Construction of the chatter detection model

The methodology of using the pre-trained neural network model is presented in Figure 8. The models we use in this approach are trained on ImageNet data containing rather generic and especially not FFT images of the vibration signal. For these models to work on a new dataset, it is necessary to proceed to a process of readjusting the models based on the feature extraction part.

This readjustment process allows in a first step to modify the outputs of these models from 1000 ImageNet categories, where they manage to find a good performance, to 3 categories (chatter, machining without chatter, and rotation without machining). The last fully connected layers used to classify 1000 ImageNet categories are removed. Secondly, all weights of the pre-trained neural network models are frozen to fit the new three-class classification task. The third step is to add three fully connected layers to perform the classification of our three classes. Following these steps, all layers of the new model are trained on the new dataset containing the FFT images from the industrial vibration signals using the softmax function to perform a multi-class classification task. Table 2 shows the hyperparameters used in the training phase.

*Table 2. Hyperparameters used in the training phase.*

| Hyperparameters | |
|---|---|
| Batch size | 2 |
| Learning rate | 0,0001 |
| Epoch | 30 |
| Optimizer | RMSprop |
| Dropout rate | 0.3 |
| Loss | Categorical Cross-Entropy loss |

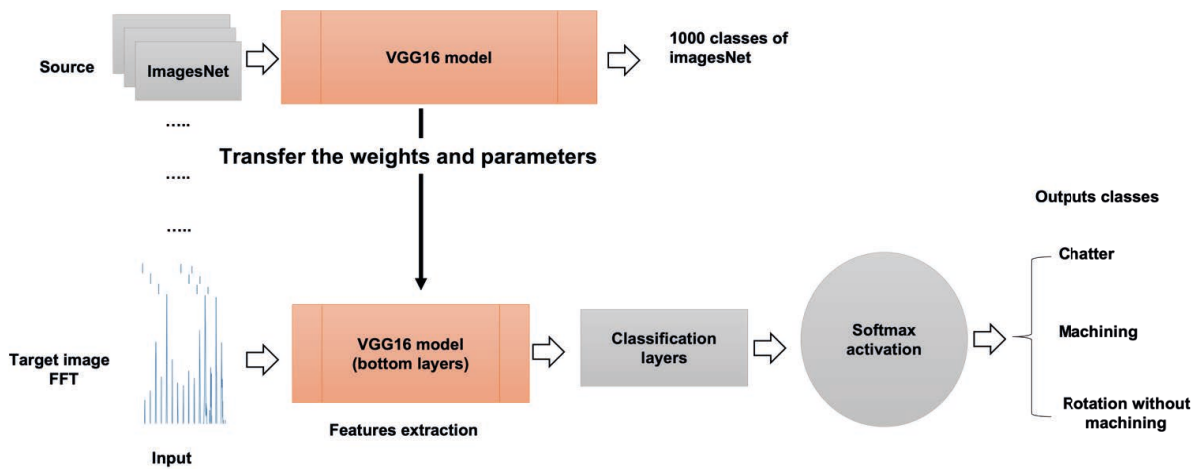

*Figure 8. The proposed transfer learning model.*

## 4. Result

This part consists in presenting the results obtained from the proposed approach. The following diagrams show our dataset's precision and loss curves based on feature extraction by VGG16 and, ResNet50. Since this study uses transfer learning techniques to extract features from the dataset. The models are trained over 30 epochs. The stabilization of the training can be seen in the curves of Figure 9 and Figure 10 from the fourth epoch.

Figure 9 shows an excellent fit case with the decrease in training and validation loss over the validation dataset up to the point of stability with the minimal deviation between training and validation loss. Figure 10 illustrates the results with ResNet50, representing inflection points in the loss and accuracy. Despite a sawtooth existence, the model has a rather satisfactory result.

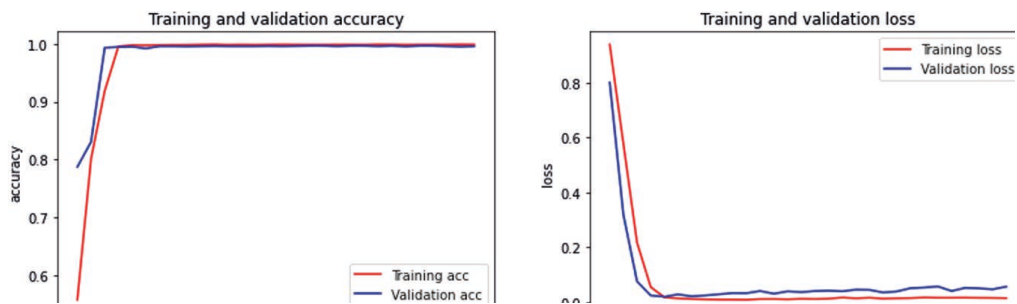

*Figure 9. Precision and loss curves with VGG16.*

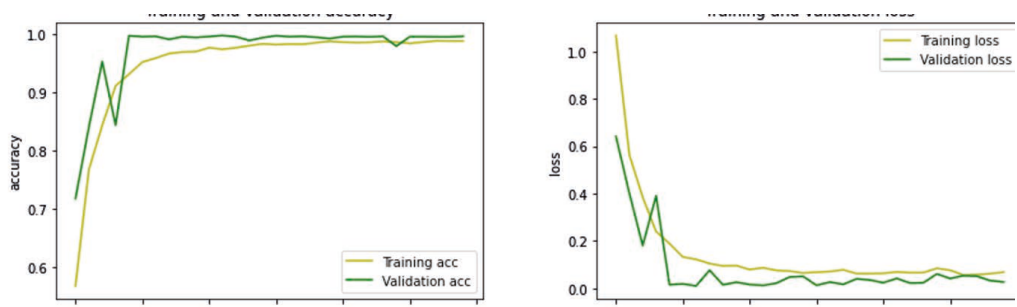

*Figure 10. Precision and loss curves with ResNet50.*

Then the model is tested with a dataset containing signals like those used in training and others, which are noisier or more ambiguous, to test our model's reliability and generalization ability. On this basis, we find an accuracy of 98.82% with the VGG16. We use techniques such as the confusion matrix and the F1 score to visualize these results to show how well our model can make interesting predictions on unknown and non-trivial data.

To get the table of classification measures, we had to convert our test data into a different format, a table that collects our test data into a single set and identifies them by several indices.

In Table 3, the most influential and essential factors are precision and the F1 score. The higher this score, the better the model used.

*Table 3. Classification report*

|  | Precision % | Recall % | F1-score % | support |
|---|---|---|---|---|
| Machining with chatter | 100 | 100 | 100 | 1544 |
| Machining without chatter | 93 | 100 | 96 | 733 |
| Rotation without machining | 100 | 98 | 99 | 2814 |

Precision is the ratio between the number of individuals correctly assigned to a class and the total number of individuals assigned to the class. Recall defines the ratio between the number of individuals correctly assigned to a class and the number of individuals belonging to the class. The F1-score combines the model's accuracy and recall, defined as the harmonic mean of the model's accuracy and recall.

To show the ability of a deep learning model to predict, we compare its result to reality by exploiting the confusion matrix. In Figure 11 and Figure 12, we can visualize the calculation of the confusion matrix on our test dataset.

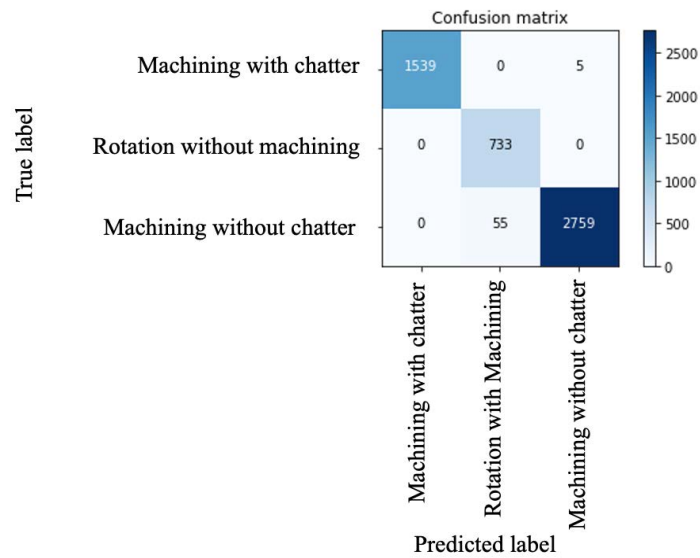

*Figure 11. Confusion matrix on unambiguous data.*

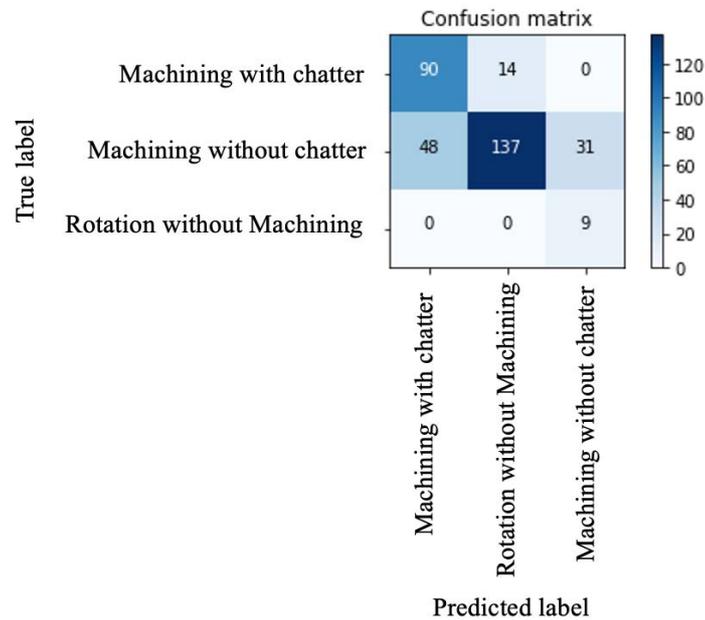

*Figure 12. Confusion matrix on ambiguous data.*

From 1544 samples of noiseless test data in the class machining with the chatter, see the first row in Figure 11 only five images were misclassified into the machining class without chatter. All images in the rotation without machining class are true positives. From 2814 images in the machining without chatter class, 55 are classified into the rotation without machining class, and the remaining 2759 are well classified. After evaluating the ambiguous data, the model obtains a prediction accuracy of 74%. It can be seen in Figure 12 that most of this discrepancy is in the class machining without chatter, of which 48 items are considered machining with chatter (the human expert has also struggled to classify this category).

On each matrix column, we find a class predicted by our model and the representations of the real classes. There are generally four categories in the matrix:

True Positive (TP): means that the prediction and the actual value are positive.

True Negative (TN): the prediction and the actual value are all negative.

False Positive (FP): the prediction is positive while the actual value is negative.

False Negative (FN): the prediction is negative while the actual value is negative.

To evaluate the output density of our three classes, we plot a curve (Figure 13) called ROC (Receiver Operating Characteristic), which calculates the ratio between the true positive rate (TPR) and the false positive rate (FPR).

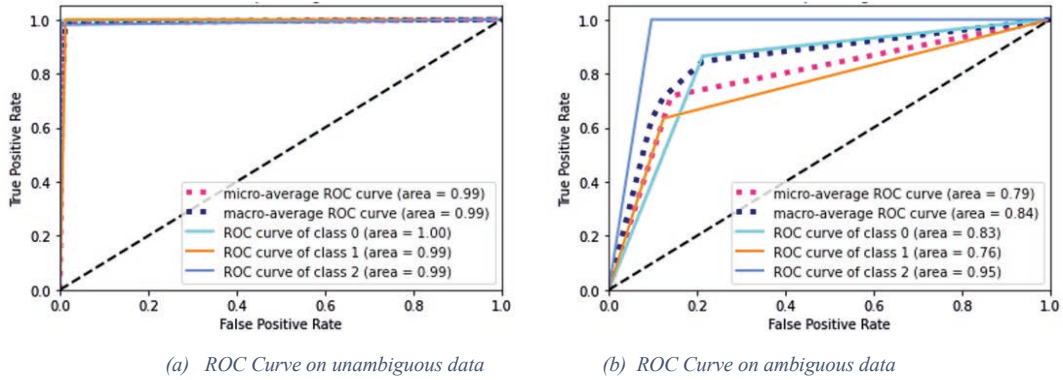

| (a) ROC Curve on unambiguous data | (b) ROC Curve on ambiguous data |

*Figure 13. ROC curve of three classifiers on ambiguous and unambiguous data.*

These curves are used to compare our different classifiers. The higher the value of a curve, the larger the area under the curve and the less error the classifier makes. In a simplified way, the classifier returns an answer that is either wrong or right. The area under the ROC curve interprets this probability. If the area value of a classifier is less than 0.5 (50%), it indicates that the marker is uninformative. An increase in the area indicates an improvement in discriminatory abilities, with a maximum of 1.0 (100%). In our case, with unambiguous data, all the values of our classifiers are close to 100%, and even in the case of ambiguous data, only one value is less than 80%. This corroborates the generalization capacity of the proposed model. Moreover, it should be noted that the labeling performed by the expert on these ambiguous data had the reliability of the same order, self-estimated by him at ≈90%, which further shows the quality of the analysis proposed by our model.

## 5. Discussions

Deep learning feeds an algorithm with data so that it learns by itself to perform a given task. In classification problems, this learning predicts results that must be compared to reality to measure its performance. The confusion matrix, also called the contingency table, is generally used. It will highlight the correct and incorrect predictions and indicate the type of errors made.

Detection of chatter using deep neural networks has become a trend in recent literature. These detection techniques in the literature use data from machining conditions far removed from industrial applications. The information on vibration signal amplitude, frequency, and rotational speed changes is strongly related to the chatter in the published data. With all this information, one wonders if all these sophisticated techniques are better than simple Root Mean Square monitoring. Since these studies use different datasets and preprocessing methods, comparing these techniques in terms of learning accuracy is difficult. We have tried to compare in Table 4 some studies concerning preprocessing methods.

In this paper, we are first interested in the types of data (industrial data or laboratory data), the preprocessing methods (FFT, STFT, CWT, etc.) of these data, and other information (amplitude, frequency, etc.) that can be linked to chatter.

For example, (Sener et al., 2021) linked chatter first to rotational speed by observing that at speeds above 8000 rpm, chatter is no longer detectable in the wavelets because the low speed did not create enough amplification. They also found that the increase in the amplitude of the signal from 10000 Hz means the appearance of the chatter phenomenon. (W. Zhu et al., 2020) propose the combination of empirical ensemble mode decomposition (EEMD) and dimensionless non-linear indicators to detect chattering. EEMD focuses on the raw signal due to its suitability for decomposing non-linear and non-stationary signals. Correlation analysis is applied to obtain the chatter-related components of the intrinsic mode function (IMF). Subsequently, the non-linear sample entropy (SE) and energy entropy (EE) of IMFs can be extracted as two indicators. (Tran et al., 2020) use CWT to transform one-dimensional time-domain signals into a two0-dimensional time-domain representation. The multi-resolution analysis performed by CWT is advantageous for analyzing non-linear and non-stationary signals. (Rahimi et al., 2021) Apply a neural network model combined with a physics-based model to detect chatter in milling. Data collected during machining are converted to short-term moving frequency spectrum using the STFT transformation process, whose features are mapped to five machining states, such as air cutting, workpiece entry and exit, stable cutting, and chatter conditions. These studies have achieved very good performance in terms of learning accuracy. However, the data collected in these studies are from laboratory vibration signals that do not represent actual industrial machining conditions. In addition, the vibration amplitude is still strongly related to the chatter in the data from these studies.

As we can see in our normalized data, our model is good enough to classify the machining type without considering the known rotation speed or amplitude of the signal, which is remarkable and distinguishes it from the bibliography. After learning and validating the data with our model, excellent results are obtained with an accuracy of 99.71% for vgg16 and 99.67% for resnet50. Table 4 presents a list of some of the works that have used artificial intelligence techniques to detect chatter with the best performance in terms of classification accuracy.

*Table 4. List of some of the best-performing works in terms of classification accuracy.*

| REF. | Pretreatment | Input Data | Classification | Precision |
|---|---|---|---|---|
| **This paper** | **FFT** | **Images FFT** | **Multilabel** | **99,71 %** |
| (Sener et al., 2021) | CWT | Images-cutting parameters | Multilabel | 99,88 % |
| (W. Zhu et al., 2020) | Size reduction | Images | Binary | 98,26 % |
| (Tran et al., 2020) | CWT | Images | Multilabel | 99,67 % |
| (Rahimi et al., 2021) | STFT | Images | Multilabel | 98,90 % |

The model needs the FFT image as input, which is very easy to generate, and some labeled data on typical signals to train the model. It takes care of itself to classify these images into their respective classes. We tested the model to see how well it can generalize the prediction obtained with unambiguous signals containing the purest information for detecting machining types and other ambiguous signals with somewhat less refined cases. Our test results on the ambiguous signal showed us that learning could be done on simple data and that recognition naturally works well on more complex data. These ambiguous signals generally make human analysis difficult, and since these signals may be brief, there is little data available to perform training on such signals. Our multiple trials before showed us that this recognition capacity is due to the power of transfer learning for feature extraction and to the fact that we have only considered the significant peaks on the generated FFT images. Compared to the approaches proposed in the literature for defect detection, always showing a trivial link between chatter and signal amplitude, we have proved that it is possible to remove the amplitude of the signal and even remove the link to a given rotation speed (because our data corresponded to several speeds of rotation used). This being said, since the network can exploit no trivial link, the question arises of the explicability of the answer given by the network.

This approach shows that it is not necessary for such research to forbid the use of neural networks pre-trained on very different images, probably as long as they remain images that humans can interpret. Also, it may be very efficient in such research to only label unambiguous signal. Then we must note that that such neural network still works fine by removing basic information such as signal amplitude, which could have been a bias in previous studies.

This model is limited by the fact that we are considering a specific machining operation, thus it would be necessary to validate the approach on other types of machining. Another limitation is the strong solicitation of a vibration expert to correctly label the data, knowing that the human expert analysis is, in fact, based on several modalities: FFT, STFT, the image of the machined surface, even sometimes forces, motor currents, and digital twin. This model also needs more explicability, as the vibration expert has no idea how the model could detect chatter with the absence of information on the amplitude and speed of rotation. Regarding the detection of chatter, human experts rely on FFTs, but also on spectrograms (STFT), to better identify the phenomena by also detecting the transitions between phenomena, which is a form of multimodality, although based on the same starting signal. The two types of transformation (FFT and SFTT) on the same signal to obtain two types of images are very different and also associated with different temporalities (Figure 14).

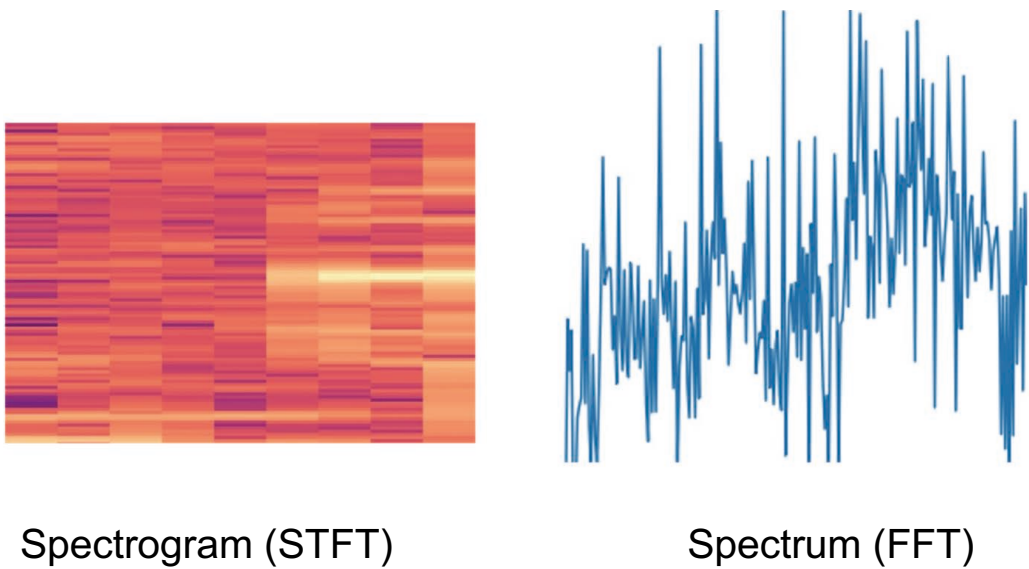

*Figure 14: Images of the sources (SFTT and FFT) of data.*

To alleviate the data labeling phase, we plan to use unsupervised learning models in which the associated model will learn to explore using multiple scales of FFT and STFT and understand data and then group them into subsets of similar elements (clustering). With unsupervised learning, the model could go so far as to tell us about a change in behavior on a machine and show us where the change comes from. It could even feed the expertise by discovering clues in the data that remain invisible to the expert's eye for the moment, as mentioned in the previous chapter, notably by methodically exploring different resolutions of FFT and STFT analysis, which the expert generally does not have the time to do. The data extracted by the different transformation techniques (FFT, STFT) could then be categorized by unsupervised techniques such as the K-nearest neighbor (KNN) or K-means algorithm. The clusters obtained by these methods are considered as classes and will serve as input data for a supervised learning model. The goal of unsupervised learning is to alleviate the process of labeling the signal by the expert.

To better understand the black-box nature of deep learning models, we plan to use explainability methods (GradCAM, lime, etc.) on the detection model prediction to visualize the operation of these deep neural networks, and pass the information to the machinist or expert. The aim is to provide information that can be understood by humans, allowing them to understand on which aspects these methods are based to detect this or that machining state in the images provided. The visual explanation method Gradient-weighted Class Activation Mapping (GradCAM) proposed by (Selvaraju et al., 2016) is an example of making decisions more transparent on CNN-based models. GradCam is applicable to any type of concept (data) introduced in a classification neural network. This technique explores the network from the convolution layer to the final layer to produce a localization map of the determining features. It highlights the important regions of the image that led to an accurate response. (Kim and Kim, 2020) used this technique for the diagnosis of rolling defects. Local Interpretable Model-Agnostic Explanations (LIME) is also one of the visualization methods that contributes to the explanation of individual predictions. This technique is independent of the prediction model and can therefore be applied to any supervised regression or classification model. LIME presented by (Ribeiro et al., 2016) is applicable to three types of data (spreadsheet, text, and images). The idea would be that after the realization of the phases of extraction of the characteristics and the classification of the images (spectrograms and spectrum) resulting from the STFT and FFT on the vibratory signal of machining, the models of visual explanations XAI are applied on the predictions to obtain a better discernment between the machining phases.

## 6. Conclusion

In this paper, the proposed approach combines vibration analysis and deep learning to detect chatter in an industrial machining process, with no-optimal sensor placement and real machining noise and uncertainties (about the real spindle). With the raw signal at the source, a powerful model was developed for detecting the different machining phases by applying the Fast Fourier Transform on the signal to generate the images that trained the model. The use of pre-trained networks using transfer learning allowed us to extract features from the data to support a generalization of the model. In the classical method of vibration analysis, several parameters are taken into account, explicitly or implicitly, to define the different phases, particularly the amplitude of the signal and the peak frequencies relative to rotation speed. In a real industrial context, this information is not always reliable and may mislead analysis; thus, it would be helpful to identify chatter without this information.

The results of the verification and visualization by the ROC curve and the confusion matrix show that the application of deep learning techniques on industrial machining data obtains excellent performances. It is essential to know that the signal's amplitude is not transmitted to the network (each FFT is recalibrated to its maximum value) nor the rotation speed (several rotation speeds were considered in the datasets). The model thus shows its capacity to identify characteristic information of the chatter in the spectrum and adapt itself to several rotation speeds, which puzzled the expert who labeled the training dataset. Moreover, some test data are similar to the training dataset, but one part is more equivocal. Again, the neural network detected these more complex cases well from its training on the easy cases.

## Acknowledgments

We want to thank Alstom Transport, in particular Mr. Antoine PALEM, for his authorization to use the data in this article.

## Ethics declaration

**a. Funding**
The authors declare that they have no known competing financial interests or personal relationships that could have appeared to influence the work reported in this paper.

**b. Conflict of Interest**
The authors of the manuscript certify that they have NO conflicts of interest to declare and that they have NO affiliation or involvement with any organization or entity with a financial interest (such as honoraria, scholarships, participation in speakers' bureaus, membership, employment, consulting, stock ownership, or other involvement; and expert testimony or patent-licensing agreements), or non-financial interest (such as personal or professional relationships, affiliations, knowledge, or beliefs) in the subject matter or materials discussed in this manuscript.

- **c. Availability of data and material (data transparency):** Not applicable.
- **d. Code availability (software application or custom code):** Not applicable.
- **e. Ethics approval (include appropriate permissions or waivers):** Not applicable.
- **f. Consent to participate (include appropriate statements):** Not applicable.
- **g. Consent for publication (include appropriate statements):** Not applicable.